\documentclass[12pt]{article}

\usepackage{amsfonts,graphicx,amsmath,amssymb,amsthm,geometry,bbm,hyperref,nicefrac,enumerate,comment,fontawesome,mathtools,cases,algorithmic,algorithm2e,stmaryrd,pict2e,picture}
	
\usepackage{color, colortbl}
\definecolor{Gray}{gray}{0.9}

\DeclareMathOperator*{\argmin}{arg\,min}

\newcommand{\I}{\mathbb{I}}
\newcommand{\R}{\mathbb{R}}
\newcommand{\N}{\mathbb{N}}
\newcommand{\E}{\mathbb{E}}

\renewcommand{\P}{\mathbb{P}}
\newcommand{\Q}{\mathbb{Q}}
\newcommand{\A}{\mathbb{A}}
\newcommand{\T}{\mathcal{T}}

\newcommand{\F}{\mathcal{F}}

\newcommand{\e}{\text{e}}
\renewcommand{\d}{\text{d}}

\usepackage{array}
\newcolumntype{P}[1]{>{\centering\arraybackslash}p{#1}}
\newcommand{\triple}{{\vert\kern-0.25ex\vert\kern-0.25ex\vert}}
\usepackage{mathtools}

\newcommand{\eqdef}{=\vcentcolon}
\usepackage{geometry}
\newtheorem{example}{Example}[section]

\theoremstyle{definition}

\newcommand{\footremember}[2]{%
    \footnote{#2}
    \newcounter{#1}
    \setcounter{#1}{\value{footnote}}%
}
\newcommand{\footrecall}[1]{%
    \footnotemark[\value{#1}]%
} 

\begin{document}

\title{D-TIPO: Deep time-inconsistent portfolio optimization with stocks and options}
\author{%
  Kristoffer Andersson\footremember{alley}{Mathematical Institute, Utrecht University, Utrecht, the Netherlands} 
  \and Cornelis W. Oosterlee\footrecall{alley}  
 }


\maketitle
\begin{abstract}
In this paper, we propose a machine learning algorithm for time-inconsistent portfolio optimization. The proposed algorithm builds upon neural network based trading schemes, in which the asset allocation at each time point is determined by a a neural network. The loss function is given by an empirical version of the objective function of the portfolio optimization problem. Moreover, various trading constraints are naturally fulfilled by choosing appropriate activation functions in the output layers of the neural networks. Besides this, our main contribution is to add options to the portfolio of risky assets and a risk-free bond and using additional neural networks to determine the amount allocated into the options as well as their strike prices.

We consider objective functions more in line with the rational preference of an investor than the classical mean-variance, apply realistic trading constraints and model the assets with a correlated jump-diffusion SDE. With an incomplete market and a more involved objective function, we show that it is beneficial to add options to the portfolio. Moreover, it is shown that adding options leads to a more constant stock allocation with less demand for drastic re-allocations.

\end{abstract}
\tableofcontents

\section{Introduction}
Mean-variance (MV) portfolio optimization, originally proposed in \cite{markowitz}, in 1955, has ever since been a cornerstone of modern portfolio selection. The popularity for practitioners as well as researchers can on the one hand be explained by its simplicity with an intuitive trade-off between reward (mean) and risk (variance) and, on the other hand, by some delicate mathematical and philosophical properties. In the original version of MV optimization, the problem was static in the sense that the allocation between a set of assets and a riskfree bond was determined at $t=0$ and held until terminal time $T$. Moreover, the assets were described by an arithmetic Brownian motion and, all together, the problem offered a closed-form allocation which was optimal with respect to the MV objective. Ever since, many extensions and generalizations have been proposed under the MV umbrella, such as static multi-period trading with discrete reallocation, see \textit{e.g.,} \cite{MP_1,MP_2,MP_3,MP_4} and continuous trading, see \textit{e.g.,}. \cite{MV_robustness,TCMV,DMV_1,DMV_2}. In most of the later applications, the arithmetic Brownian motion has been replaced by geometric Brownian motion or by more general processes describing the asset dynamics. 
\subsubsection*{Time-inconsistency}
Despite the straightforward formulation of the MV optimization, there are several difficulties making standard approximation tools from stochastic optimal control theory intractable. The main problem is that the MV objective function does not satisfy the law of iterated expectations and hence, the dynamic programming principle (DPP) is not satisfied. From an intuitive perspective, this means that a strategy which is optimal at time $t$ is in general not optimal at $t+h$. From a practical perspective, this implies that the standard toolbox of computational methods for problems satisfying the DPP (i.e., optimization and approximations of conditional expectations backward in time) is not applicable. There are many attempts to circumvent this issue and they can mainly be categorised into two classes; \textit{i}) \textit{Pre-commitment strategies}, in which an embedding technique is used to transform the problem into an equivalent problem which satisfies the DPP, see \textit{e.g.,} \cite{DMVO_LQf}, or \textit{ii}) \textit{Time-consistent strategies} in which optimization is performed over the subset of time-consistent allocation strategies. Both methods have their own advantages and disadvantages and can both be motivated from different philosophical perspectives. For a more detailed discussion, we refer to \cite{Vinga1, Vinga2, Forsyth1}. 

The two classes above have in common that they transform the problem into a time-consistent problem, in which the DPP holds. In turn, this implies that the problem can be solved recursively, backward in time, one sub-problem at the time. By solving many sub-problems, classical methods to approximate conditional expectations can be used. Typically, one can resort to Monte--Carlo methods or approximation methods for PDEs or FBSDEs, see \textit{e.g.,} \cite{MP_1,DMV_2} and \cite{PDE_1, PDE_2, ji2020three, han2020convergence, paper3}, respectively. Even though there are certain ways to reformulate a MV optimization problem either using a probabilistic approach, to obtain a McKean--Vlasov FBSDE or a PDE approach to obtain a so-called master equation, see \textit{e.g.,} \cite{MVFBSDE}. These problems are also inherently difficult to (numerically) solve. 

In this paper, instead of a problem specific embedding technique, we follow \textit{e.g.,} \cite{pre_1, Forsyth1, Forsyth2, pre_2} and employ a machine learning algorithm to approximate the allocation strategy. Our approach is to take a step back and view the portfolio optimization problem as a stochastic control problem. After a discretization in time, the optimization problem can be approximated by representing the allocation strategy with a sequence of neural networks and letting the loss function be an empirical version of the objective function. This was originally proposed in \cite{han2016deep} to approximate general stochastic control problems and in \textit{e.g.,} \cite{Forsyth1,Forsyth2,MV_robustness} specifically for portfolio optimization. These algorithms have a clear advantage for time-inconsistent problems since the optimization is performed only once. Moreover, in contrast to the classical dynamic programming approach, the problem is solved forward in time, and as a consequence, the algorithm does not rely on the DPP and hence, time-consistent and time-inconsistent problems are treated similarly.

\subsubsection*{Adding options and gain flexibility}
Due to the ambiguity of measuring risk with variance of the terminal wealth, many alternative ways to measure risk have been proposed. As the most common alternatives, we mention the semi-variance and expected shortfall, which are explored in \textit{e.g.,} \cite{MSV_1,MSV_2,MES_1,MES_2}. As stated above, it is clear that the MV objective function is not completely in line with the rational preference of an investor, since it treats downside risk and upside potential equivalently. On the other hand, it is not clear whether optimizing with respect to another objective function will result in strategies that are more in line with the investors' preferences. As an example, assume that the asset returns are normally distributed, then the MV objective is, objectively, the best objective function since the law of the normal distribution is completely determined by its mean and variance. The question is then whether or not asset returns are normally distributed, and the answer is, in general, a clear ''no''. On the other hand, asset returns are usually close to symmetric and a symmetric distribution with reasonable tails can be relatively well approximated by a normal distribution. Therefore, it is not clear that we can create a terminal wealth with a distribution that is flexible enough to benefit from the more complex objective functions. Fortunately, there are several financial products with asymmetric returns, which can be added to a portfolio. For instance, a plain European option has a highly asymmetric return and is a good candidate to add to the portfolio. Although a straightforward extension to a stock and bond portfolio, portfolio optimization with equity options has not been 
a major topic in the scientific literature. Some attempts are found in \cite{PO_w_options,options_1,options_2,options_3,options_4}, but the focus was mainly on static portfolios consisting of only options. The focus in this paper is, instead, to combine the two approaches and trade in both the assets themselves and European options written on the same assets, and have a machine learning algorithm determine the optimal allocation. We name the proposed algorithm \emph{Deep Time-Inconsistent Portfolio optimization with stocks and Options}, with acronym ''D-TIPO''. When the options are not included in the allocation learning, we use the abbreviation ''D-TIP''.

\subsubsection*{Structure of the paper}
This paper is organized as follows. In Section \ref{sec2}, the framework is introduced. We outline trading constraints, objective functions and aim to provide some heuristic motivation for the objective functions used. In Section \ref{sec3}, the methodology including discretization of the continuous optimization problem, as well as the neural networks used, are outlined. Moreover, we explain how the trading constraints can be built into the neural network structure and provide pseudo-code for the full algorithm. Section \ref{num_exp} contains numerical experiments in which we first confirm the accuracy of the algorithm for an example with a reference solution and then the algorithm is validated and compared with some other allocation strategies. Finally, in section \ref{conclusions}, we conclude about the findings of the paper.

\section{Problem formulation}\label{sec2}
In this paper, we take the perspective of a \emph{trader}, who is allowed to trade in one risk-free bond, in $N^\text{stocks}\in\N$ stocks, and $N^\text{options}\in\N$ options. We let $S=(S_t)_{t\in[0,T]}$ be an $\R^{N^\text{stocks}}-$valued time-continuous Markov process on a complete probability space $(\Omega,\,\mathcal{F},\,\A)$. We consider a \emph{trading period} $\T\eqdef[0,T]$, where $T\in\R_+$ is referred to as the \emph{terminal time}. The outcome set $\Omega$ is the set of all possible realizations of the stochastic economy, $\F$ is a $\sigma-$algebra on $\Omega$ and we define $\F_t$ as the sub-$\sigma$-algebra generated by $(S_s)_{s\in[0,t]}$. The probability measure $\A$ is a generic notation, representing either the real world measure, or the risk neutral measure, denoted $\P$ and $\Q$, respectively. The bond is denoted by $B=(B_t)_{t\in[0,T]}$ and for $i\in\{1,2,\ldots,N^\text{options}\}$, an option with $S$ as underlying asset (could be a single stock or a basket of stocks), at time $t\in[0,T]$ and terminating at $T$ is denoted by $V^i(t, S_t\,;K)$, respectively, where $K\in\R$ is the strike price. Without loss of generality, we set the initial values of all stocks, options and the bond to unity at $t=0$, \textit{i.e.,} for $j\in\{1,2,\ldots,N^\text{stocks}\}$ and $i\in\{1,2,\ldots,N^\text{options}\}$, we set $S_0^j=1$, $V^i(0,S_0)=1$ and $B_0=1$.

The trader is allowed to trade the stocks and the bond at a set of \emph{trading dates}, denoted by $\T\subseteq [0,T]$. If $\T= [0,T]$, then trading is allowed continuously in the entire trading period but a more realistic scenario is that trading is allowed only on a discrete set of dates within the trading period. The options, on the other hand, can only be traded at \emph{initial time} of the trading period, \textit{i.e.,} at time 0 and are held until the terminal time. It should however be pointed out that our framework would also allow for trading in the options during the entire time period $\T$, but since options are less liquid and transaction costs usually are significantly higher than for stocks, we restrict ourselves to trade options only at $t=0$. 

Denote by $\alpha^k=(\alpha_t^k)_{t\in[0,T]}$ the piecewise constant process describing the amount the trader holds in stock $k$ and for $k=0$, the amount the trader holds in the bond. We can describe the total wealth of the portfolio stemming from the stock and the bond holdings at time $t$ by
\begin{equation}\label{x_t_simple}
   x_t=\alpha_t^0B_t + \sum_{k=1}^{N^\text{stocks}}\alpha_t^kS_t^k.
\end{equation}
Since the portfolio is self-financing, \begin{equation}\label{alpha_0}
    \alpha_t^0=\frac{1}{B_{\tau(t)}}\bigg(x_{\tau(t)} -   \sum_{k=1}^{N^\text{stocks}}\alpha_{\tau(t)}^kS_{\tau(t)}^k\bigg),
\end{equation}
where $\tau(t)=\max_s\{s\in\T\,|\,s\leq t\}$, \textit{i.e.,} the most recent trading date. Sometimes, it is convenient to work with the total amount of cash.
We therefore denote by $A^k=(A_t^k)_{t\in[0,T]}$, the amount of cash invested in asset $k$ and define for $k>0$, $A^k_t=\alpha_t^kS_t^k$ and for $k=0$, $A_t^0=\alpha_t^0B_t^0$. The return on investment from trading in the stocks and the bond is then given by \begin{equation}\label{R_S}
    R_\text{SB}(S;\alpha) = x_T - x_0.
\end{equation}

Let $\beta^i$ denote the amount of option $i$ in the portfolio. The total amount invested in the options at time $t$ is then given by
\begin{equation*}
    y_t = \sum_{i=1}^{N^\text{options}}\beta^i V^i(t,S_t;K^i),
\end{equation*}
where $K^i$ is the strike price for option $i$. We can define the return on the investment from the static option position by
\begin{equation}\label{R_opt}
    R_\text{O}(S;\beta) = y_T-y_0.
\end{equation}
Summing up \eqref{R_S} and \eqref{R_opt}, we obtain the total return \begin{equation}\label{R}
    R(S;\alpha,\beta) = R_\text{SB}(S;\alpha) + R_\text{O}(S;\beta).
\end{equation}
In order to evaluate the satisfaction of the investor with the return, we use an objective function. A good objective function should be able to numerically represent the investors view of risk. In this section, the only restriction we put on the objective function is that it is a deterministic function, taking the trading strategies $\alpha$ and $\beta$ as inputs. The objective function is of the form \begin{equation}\label{objective}
U(\alpha,\beta) = u\big(\mathcal{L}[R(S;\alpha,\beta)]\big),    
\end{equation}
where $\mathcal{L}(\cdot)$ denotes the probabilistic law. Our objective is then to find a trading strategy $\alpha,\beta$, such that the objective function is maximized. Note that we do not specify the trading strategies that are allowed. Moreover, we assume that the maximum of the objective function above is attainable.

\begin{example} Suppose we consider a portfolio consisting of only stocks, modeled with a multi-dimensional Geometric Brownian Motion and a bond with deterministic interest rate and that continuous trading on $[0,T]$ is allowed. If the aim is to maximize the expectation and minimize the variance of $R_\emph{SB}(S;\alpha)$, \textit{i.e.,} using an objective function of the form $U^\emph{DMVO}(\alpha;\lambda)=\E[R_\emph{SB}(S;\alpha)] - \lambda \text{Var}[R_\emph{SB}(S;\alpha)]$ ($\lambda\in\R_+$ is a risk parameter), we are in the classical \emph{Dynamic Mean-Variance Optimization} framework. It is well-known, see \textit{e.g.,} \cite{DMVO_LQf}, that there is an analytic strategy $\alpha^*$ such that
\begin{equation*}
    U^\emph{DMVO}(\alpha^*;\lambda)=\emph{Max}_\alpha U^\emph{DMVO}(\alpha;\lambda).
\end{equation*}
Now, if we would also allow for trading in vanilla call and put options on each of the underlying stocks, this would not increase the utility, since in the Black--Scholes framework, the options would be fully replicable by the stocks and the bond. This implies that any strategy involving stocks, the bond and options can be replicated by a strategy only involving stocks and the bond.   
\end{example}

The example above indicates that adding options to a specific portfolio does not increase utility for the trader, but is that statement true in general? The answer is that if we are able to trade continuously in a complete market, then the European option is replicable, and hence it is not possible to increase the utility by adding options to the portfolio. On the other hand, if the market is incomplete, or when we are not able to trade continuously, then there is a possibility that adding options to a portfolio can increase the utility. It is a well-known fact that the real world financial market is incomplete, and in the subsections below, we introduce some aspects which make the model of the market incomplete. It should however be pointed out that in general, a L\'evy process modelling the stocks, (\textit{e.g.,} Geometric Brownian Motion with jumps) would lead to an incomplete market, even without adding any other market frictions.

\subsection{Market frictions}
In this section, we introduce the market frictions and trading constraints considered in this paper. Below, we extend the framework to allow for transaction costs, non-bankruptcy constraint and leverage constraints for the trading strategies $\alpha$ and $\beta$. 

From now on, unless otherwise is stated, we assume that $\T$ is a set of finite trading dates\footnote{Or a discrete approximation of continuous trading. In practice, all trading is carried out in a time discrete fashion, but in some cases a closed-form optimal strategy can be achieved in the continuous case, which is why we allow for such trading in our framework.}, \textit{i.e.,} $\T=\{t_0,t_1,\ldots,t_{N-1}\}$, such that $t_0=0$ and $t_i<t_{i+1}$. We can now rewrite the value of the stock and the bond, given in \eqref{x_t_simple}, in terms the initial value plus the accumulated increments between the finite trading dates, i.e.,
\begin{equation}\label{x_n}
    x_{t_{n+1}} = x_{t_n} + \alpha_{t_n}^0 (B_{t_{n+1}} - B_{t_n}) + \sum_{k=1}^{N^\text{stocks}}\alpha_{t_n}^k(S_{t_{n+1}}^k - S_{t_n}^k).
\end{equation}
\newline\newline
\noindent\textbf{Transaction costs:} Assuming a proportional transaction cost we can write the sum of the transaction costs for stock $k$ as
\begin{equation}\label{TC}
\text{TC}^k = \sum_{n=1}^{N}C\e^{r(T-t_n)}(\alpha^k_{t_n}-\alpha^k_{t_{n-1}})S_{t_n}^k.
\end{equation}
Here $100\times C\in\R_+$ represents the transaction costs as a percentage of the size of the transaction. Finally, the total transaction cost is given by \begin{equation*}
    \text{TC} = \sum_{k=1}^{N^\text{stocks}}\text{TC}^k.
\end{equation*}
In fact, the implication of the above is that we do not pay transaction costs immediately, but instead at the end of the trading period, with appropriate interest rate. 
\newline\newline
\noindent\textbf{Non-bankruptcy constraint:} 
The non-bankruptcy constraint is formulated such that the first time the sum of the values of the stocks and the bond is non-positive, the portfolio is liquidated. This implies that, in the presence of a non-bankruptcy constraint, \eqref{x_n} is replaced by
\begin{equation}\label{x_n_NBC}
    x_{t_{n+1}} =  x_{t_n} + \I_{\{x>0\}}(x_{t_n})\bigg(\alpha_{t_n}^0 (B_{t_{n+1}} - B_{t_n}) + \sum_{k=1}^{N^\text{stocks}}\alpha_{t_n}^k(S_{t_{n+1}}^k - S_{t_n}^k)\bigg),
\end{equation}
where $\I_{\{x>0\}}(\cdot)$ is the indicator function. 

The above implies that a liquidation of $x$ caused by a bankruptcy can only occur at a trading date. In reality, it would be reasonable to liquidate the portfolio as soon as the value touches zero. On the other hand, immediate liquidation may not be possible, due to low liquidity. It should also be noted that the underlying methodology does not rely on this specific modelling choice.  
\newline\newline
\noindent\textbf{Trading constraint:} 
There are several different trading constraints and below, we mention a few of them. \begin{itemize}
    \item \textbf{No short-selling constraint} - No short-selling of the stocks, implying that for $t\in[0,T]$ and $1\geq k\geq N^\text{stocks}$, $\alpha_t^k\geq 0
    $.
    \item \textbf{No leverage constraint} - Implying that we cannot short sell the bond, \textit{i.e.,} for $t\in[0,T]$, $\alpha_t^0\geq 0$.
    \item \textbf{No bankruptcy constraint} - If $x_{t_n}\leq 0$, then all positions are liquidated and for $t\geq t_n$, $x_t=x_{t_n}$.
\end{itemize}
The aim in this paper is to apply a neural network based strategy to approximate the allocations at each trading date. The constraints described above are built into the neural networks, which is specified in full detail in sections below.

\subsection{Objective functions}
There are several different ways of measuring the performance of a trading strategy. What all measures have in common is that they attempt to numerically represent the risk aversion of the trader so that a strategy optimal for the trader's preferences can be adopted. Below, we introduce a few different objective functions and present their individual motivations.
\newline\newline
\textbf{Utility functions:} A utility function measures the so-called marginal happiness of wealth, \textit{i.e.,} how happy is the investor with one extra unit of wealth, given the current level of wealth. These functions are typically concave since one extra unit adds more happiness when the investor's wealth is low than when it is high. The objective function used is usually the expected utility, which leads to a mathematically nice framework from a modeling perspective since the problem becomes time-consistent and, hence, the dynamic programming principle holds. A disadvantage with these types of objective functions is that they are not particularly intuitive and it becomes therefore difficult for managers to determine the risk parameters. Another disadvantage is that the expected utility is a narrow measure of the utility and does not say anything about the variability of the utility outcomes.   
\newline\newline
\textbf{Mean-Variance objective:} Under the assumption that asset returns are normally distributed, the mean-variance objective function is the optimal choice, since the normal distribution is completely determined by its mean and variance. It is however a well-known fact that asset returns are not normally distributed, but have a slightly fatter tail than the one of a normal distribution. This phenomenon is even more pronounced for volatility strategies (strategies implemented in the derivatives market using options and other financial products), see \textit{e.g.,} \cite[page 40]{factor_investing}, in which these two strategies are compared in terms of their first four moments. It is shown that while the first two moments are almost identical, the third and fourth moments differ significantly (the third and fourth moments for asset returns are closer to being normal while these moments for the returns of a volatility strategy exhibit significant leptokurtic behaviour). The numbers are displayed in Table \ref{strategies}. 
\begin{table}[] \centering
\begin{tabular}{lll}
                   & \textbf{Volatility strategy} & \textbf{Equities} \\ \hline
Mean               & 9.9\%                        & 9.7\%             \\
Standard deviation & 15.2\%                       & 15.1\%            \\
Skewness           & -8.3                         & -0.6              \\
Kurtosis           & 104.4                        & 4.0              
\end{tabular}
\caption{Comparison of returns of a volatility strategy and a pure equity strategy, in terms of mean, standard deviation, skewness and kurtosis. If returns would have been normally distributed, the skewness and the kurtosis would have been 0 and 3, respectively.}
\label{strategies}
\end{table}
It should be emphasized that by using the mean-variance objective function, we do not inherently assume normally distributed returns but optimality in terms of the trading strategy related to a mean-variance objective function could be ambiguous as can be demonstrated with the example above. Since the mean values (9.9\% and 9.7\%) and the variances (15.2\% and 15.1\%) are similar in a mean-variance objective function, one could conclude that the strategies are equally good, even though the distributions of the returns are very different. The question is then which of the two strategies a trader should apply? If the trader manages a pension fund, it is reasonable to believe that a high negative skewness and a large kurtosis are highly undesirable, while for a hedge fund with short time horizon the reasoning could be the other way around. The challenge for us is then to construct an objective function which better represents the true objective compared to the mean-variance objective. 

\begin{figure}[htp]
\centering
\begin{tabular}{c}
          \includegraphics[width=130mm]{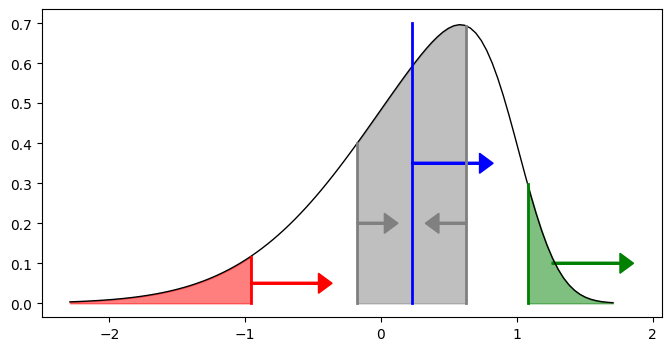}
          \end{tabular}
\caption{Example of a probability density function for the terminal wealth and the terms in \eqref{MV_ES_ES_objective} are illustrated. Red, blue and green represent the lower expected shortfall, mean and higher expected shortfall, respectively, and the arrows represent the preferred direction. 
 The gray area represents the mean plus/minus the variance and the arrows represent the preference to decrease variance.}\label{MV_ES_ES}
\end{figure}

As a final note on when it makes sense to consider another objective, we claim that when the returns deviate from normality and in particular if the distribution is asymmetric, it is natural to ask why a trader would want to penalize not only downside risk but also the upside potential, which is the direct consequence of using the variance as a measure of risk.
\newline\newline
\textbf{Non-symmetric objectives:}
Since we deal with trading strategies that are able to create highly non-symmetrical distributions of the terminal wealth, we want to explore objective functions suitable for this. The first important aspect is to model downside risk and upside potential differently. One way to assess this problem is to use \textit{expected shortfall}, which is defined as the mean of the tail of a distribution. For example, if we want to penalize downside risk, we may try to maximize the average of the 10\% worst outcomes or if we want to encourage upside potential, we could try to maximize the average of the 10\% best outcomes. The expected shortfall can be defined using the \emph{Value at Risk} in the following way
\begin{align*}
    \text{VaR}_p(R) &= \inf\{P\in\R\,|\,\A(R\leq P)\geq p\},\\
    \text{ES}^-_p(R) &= \E[R\,|\,R\leq \text{VaR}_p(R)],\quad \text{ES}^+_p(R) = \E[R\,|\,R\geq \text{VaR}_p(R)].
\end{align*}
For notational convenience, we omit the dependency on the trading strategy in the notation below. A typical objective function would then be 
\begin{equation}\label{MV_ES_ES_objective}
   U =u\big(\mathcal{L}(S)\big) = \E[R] - \lambda_1 \text{Var}[R] + \lambda_2\text{ES}^-_{p_1}(R) + \lambda_3\text{ES}^+_{p_2}(R),
\end{equation}
where $\lambda_1,\lambda_2,\lambda_3\in\R_+$ are parameters describing the risk preference and $p_1,p_2\in(0,1)$ are the parameters controlling the sizes of the left and right tails we want to address. For instance, one could choose $p_1=0.1$ and $p_2=0.9$ to control the upper and lower tails (10\% on each side). In the formulation above, we view $R$ as a generic return of some strategy, however, in our case $U$ and $R$ would depend on the trading strategy and $R$ would in addition also depend on a random realization of the stochastic economy, as in \eqref{R} and \eqref{objective}. In Figure \ref{MV_ES_ES}, the objective function given in \eqref{MV_ES_ES_objective} is illustrated for a toy example with a returns following a skew-normal distribution. 

\subsection{Full optimization problem}
The aim in the previous sections was to explain and motivate each component of the problem statement. Here, we summarize the above and formulate the optimization problem. 

In the sections above, we referred to $\alpha$ and $\beta$ as the trading strategies. From here on, we also consider the set of strike prices, $K=(K^1,K^2,\ldots,K^\text{options})$, as part of the trading strategy which is denoted by $\pi=(\alpha,\beta,K)$. The following example 
 aims to motivate why we also include strike prices. \begin{example}[Bull-call-spread strategy]
 Consider a situation where a trader believes that a stock will either increase in value, or drastically decrease, for instance, when the company is about to report. Then, a so-called bull-call-spread can be used. This is the position where the trader both buys and sells a European call option but with different strike prices. If executed correctly, this can create a situation where the trader only risks to loose the difference in the premiums between the bought and sold option with an upside potential proportional to the stock value but bounded from above. This strategy is arguably best explained with a specific example and illustrative figures. Therefore, we consider a stock which today, at $t=0$, has value $S_0=1.0$. Moreover, we have two European call options with values at maturity $T$, $V^1(T,S_T;K^1)=\text{max}(S_T-K^1,0)$ and $V^2(T,S_T;K^2)=\text{max}(S_T-K^2,0)$, respectively. We set $K^1=1.1$ and $K^2=1.3$ and in Figure \ref{BCS} (left-side), we display the return, at maturity $T$, for the stock itself, buying one unit of option 1 and selling one unit of option 2, respectively. In the right-side figure, we compare returns of three different strategies and it becomes clear that the return is dependent not only on how much of each of the three products the trader invests in, but also on the strike prices $K^1$ and $K^2$.  
\begin{figure}[htp]
\centering
\begin{tabular}{cc}
          \includegraphics[width=80mm]{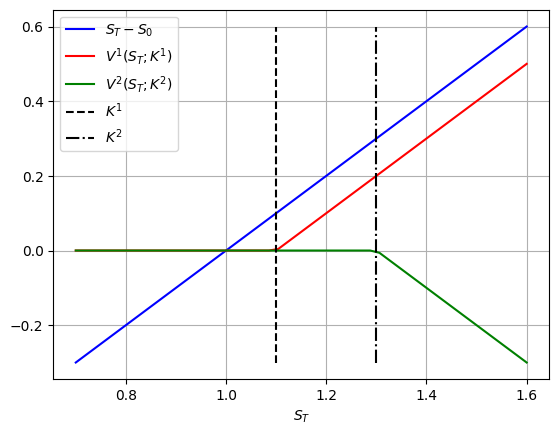}&    
          \includegraphics[width=80mm]{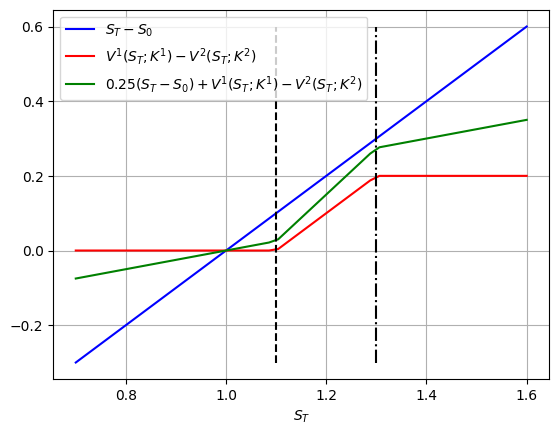}
\end{tabular}
\caption{Returns plotted against the stock value at the terminal time $T$. \textbf{Left:} Returns for investing in a stock, buying one unit of option 1 and selling one unit of option 2, respectively. \textbf{Right:} Returns for three different combinations of the products, where the red line is the classical bull-call spread.}\label{BCS}
\end{figure}
 \end{example}
 As becomes clear in the example above, it is not easy to have an intuitive overview on which strike prices to choose, especially not when a complex objective function is used. Therefore, we consider also the strike prices to be part of the optimization problem.

We assume an objective function $U(\pi)=u\big(\mathcal{L}[R(S;\pi)]\big)$, an initial wealth $x_0^\text{IC}\in\R_+$ (usually set to 1) and we denote by $\Pi$ the set of \emph{admissible} (allowed) trading strategies, \textit{i.e.} the set of executable strategies taking all the trading constraints into account.  All equations below are collected from the sections above, but the dependency on the trading strategy $\pi$ is specified more carefully to make the optimization problem more clear.
\begin{equation}\label{full_opt_prob}
\begin{dcases}
\underset{\pi\in\Pi}{\mathrm{maximize}}
 = U(\pi)=u\big(\mathcal{L}(R(S\,;\pi))\big),\quad \text{where},\\
R(S\,;\pi) = R_{\text{SB}}(S\,;\pi) + R_\text{O}(S\,;\pi)\\
R_{\text{SB}}(S\,;\pi)= x_T(S;\pi) - x_0(\pi) - \sum_{k=1}^{N^\text{stocks}}\text{TC}^k,\quad R_\text{O}(S\,;\pi)=y_T(S\,;\pi)-y_0(\pi),\\
x_T(S;\pi) = x_0 + \sum_{n=0}^{N}\I_{\{x>0\}}\big(x_{t_n}(S;\pi)\big)\big[\alpha_{t_n}^0(B_{t_{n+1}}-B_{t_n}) + \sum_{k=1}^{N^\text{stocks}}\alpha_{t_n}^k(S_{t_{n+1}}^k - S_{t_n}^k)\big],\\
\quad x_0=x_0^\text{IC}-y_0(\pi),\quad 
y_T(S\,;\pi) = \sum_{i=1}^{N^\text{options}}\beta^iV^i(T,S_T\,;\,K^i),\quad y_0(\pi) = \sum_{i=1}^{N^\text{options}}\beta^iV^i(0,S_0\,;\,K^i).
\end{dcases}
\end{equation}
This optimization problem admits to closed-form strategies $\alpha$ and $\beta$ (for fixed $K$) only in rare special cases and we therefore have to rely on numerical approximations. When it is required, we use a time discretization scheme for $S$ and we use empirical distributions as approximations for $\mathcal{L}[R(S;\pi)]$ in a Monte--Carlo fashion. The discrete counterpart (in time and/or in probability space) of \eqref{full_opt_prob} is then approximated by letting deep neural networks represent the trading strategies and optimizing with a gradient decent type algorithm. This is explained in more detail in sections below.

\section{Methodology}\label{sec3}
As briefly mentioned in the previous section, there are two entities that often need to be discretized, namely the asset price process and the objective function. For the asset process, we need to resort to a discretization scheme when there are no closed-form expressions available. The objective function usually not only depends on the returns but also on the probability distribution of the returns. Since this is unknown, despite for rare special cases, we need an approximation. This is simply done by using empirical distributions as approximations.
\subsection{Discretized asset process and empirical distribution}\label{3.1}
Let $\T^N=\{t_0,t_1,\ldots,t_{N-1}\}$ be the set of distinct trading dates, where either, $\T^N=\T$, or, if $\T$ is an infinite set (continuous trading), $\T^N$ is a discrete approximation of $\T$. We let $t_N=T$ and for $0\leq i\leq N-1$, $t_{i}<t_{i+1}$. We assume that we are able to generate $M\in\N_+$ samples of the $N^\text{stocks}-$dimensional asset process $S$. We denote asset $k$, realization $m$, at time $t_n\in\T_N$ by $S_{t_n}^k(m)$, realization $m$ at time $t_n\in\T_N$ by $S_{t_n}(m)=\{S_{t_n}^1(m),\ldots S_{t_n}^{N^\text{stocks}}(m)\}$ and realization $m$ by $S(m)=\{S_{t_0}(m),\ldots S_{t_N}(m)\}$. 

\begin{example}[A one-dimensional Geometric Brownian motion]
Given the initial value $S_0$ and assuming we have access to the drift and diffusion parameters $\mu$ and $\sigma$, a closed-form solution for the asset process is given by
    \begin{equation*}
        S_t = S_0\e^{(\mu-\frac{\sigma^2}{2})t + \sigma (W_t-W_0)}.
    \end{equation*}
A model like this can easily be sampled from for any $t\in\T^N$ and by the Markovianity of $S_t$ no information is lost by only sampling discrete values.
\end{example}
\begin{example}[A general jump-diffusion SDE]
Given the initial value $S_0$ and assuming we have access to the drift, diffusion and jump coefficients $\mu$, $\sigma$ and $J$, we employ, 
    \begin{equation*}
        \emph{d} S_t = \mu(t,S_t)\emph{d} t + \sigma(t,S_t)\emph{d} W_t + J(t,S_t)\emph{d} X_t,
    \end{equation*}
where $X_t$ represents a jump process which we do not further specify in this example. In general, we do not have a closed-form solution for $S_t$ and therefore we need to rely on a discrete approximation scheme such as the Euler--Maruyama scheme. Similar to the Geometric Brownian motion, the jump-diffusion SDE above is Markovian, but one needs to be careful with the time discretization in order to ensure convergence.
\end{example}

We denote by $\bar{U}^M(\pi)$ the objective function when the distributions of the returns are approximated by an empirical counterpart with $M$ samples and by $\mathcal{L}^M[\,\cdot\,]$ the empirical probability law.

\subsection{Neural network approximation}\label{NN_spec}
In this section, we describe how the trading strategy $\pi$ is represented by a sequence of neural networks and simple parameters. Optimality in terms of the objective function is then sought with a gradient descent type algorithm. We sometimes use the machine learning terminology \emph{loss function} and \emph{training} for the objective function and optimization procedure, respectively.
\subsubsection{General notation for a neural network}\label{FCNN}
Below, we introduce some machine learning notation, which is subsequently put into the context of this paper. A neural network, which is nothing but a parametrized mapping, is here denoted by $\phi(\,\cdot\,;\theta^\text{NN})\colon\R^{\mathfrak{D}^{\text{input}}}\to\R^{\mathfrak{D}^{\text{output}}}$, where $\mathfrak{D}^{\text{input}}$ and $\mathfrak{D}^{\text{output}}$ are the input and output dimensions, respectively. Here, $\theta^\text{NN}$ is a parameter containing all the trainable (optimizeable) parameters of the network. 
\begin{itemize}
    \item Denote the number of layers (input and output layers included) by $\mathfrak{L}\in\N$, and for layer $\ell\in\{1,2,\ldots,\mathfrak{L}\}$, the number of nodes by $\mathfrak{N}_{\ell}\in\N$. Note that $\mathfrak{N}_1=\mathfrak{D}^{\text{input}}$;
    \item For layer $\ell\in\{2,3,\ldots,\mathfrak{L}\}$, we denote the weight matrix, acting between layers $\ell-1$ and $\ell$, by $w_\ell\in\R^{\mathfrak{N}_{\ell-1}\times\mathfrak{N}_\ell}$, and the bias vector by $b_\ell\in\R^{\ell}$;
    \item For layer $\ell\in\{2,3,\ldots,\mathfrak{L}\}$, we denote the (scalar) activation function by $a_\ell\colon\R\to\R$ and the vector activation function by $\boldsymbol{a}_\ell\colon\R^{\mathfrak{N}_\ell}\to\R^{\mathfrak{N}_\ell}$, which, for $x=(x_1,x_2,\ldots,x_{\mathfrak{N}_{\ell}})$, is defined by 
    \begin{equation*}
        \boldsymbol{a}_\ell(x)=\begin{pmatrix}a_\ell(x_1)\\
        \vdots\\
        a_{\ell}(x_{\mathfrak{N}_\ell})\end{pmatrix};
    \end{equation*}
    \item The output of the network should obey the trading constraints and this is managed by choosing an appropriate activation function in the output layer.
\end{itemize}
The neural network is then defined by \begin{equation}
    \phi(\cdot\,;\,\theta) = L_{\mathfrak{L}}\circ L_{\mathfrak{L}-1}\circ\cdots\circ L_1(\cdot),\label{F_NN}
\end{equation}
where for $x\in\R^{\mathfrak{L}_{\ell-1}}$, the layers are defined as 
\begin{equation*}
L_{\ell}(x)=\begin{cases}x,&\text{for }\ell=1,\\ 
    \boldsymbol{a}_{\ell}(w_{\ell}^\top x+b_{\ell}),&\text{for }\ell\geq 2,\end{cases}
\end{equation*}
with $w_{\ell}^\top$ the matrix transpose of $w_{\ell}$. The trainable parameters are then given by the list\begin{equation*}
    \theta^{\text{NN}}=\left\{w_2,b_2,w_3,b_3,\ldots,w_\mathfrak{L}, b_\mathfrak{L}\right\}.
    \end{equation*}
Finally, denote by $\mathfrak{D}^{\theta^{\text{NN}}}$ the number of trainable parameters, \textit{i.e.,} $\mathfrak{D}^{\theta^{\text{NN}}}=\sum_{k=2}^{\mathfrak{L}}\text{dim}(w_k)+\text{dim}(b_k)$.
\subsubsection{Neural networks representing the trading strategy}
As mentioned above, the trading strategy $\pi$ consists of three parts; \emph{i}) the static amount invested in each option $\beta$, \emph{ii}) the static strike prices of the options $K$, and \emph{iii}) the dynamic amount invested in each stock $\alpha$. Recall that $\beta$ and $K$ are decided at $t=0$, and, as long as we have a deterministic initial wealth $x_0^\text{IC}$, we want to have a deterministic representation for $\beta$ and $K$ (this is what is meant by a static strategy above). The third component, $\alpha$, on the other hand, does not have to be deterministic since it may depend on previous performance, which is affected by randomness through the stock process (this is what is meant by a dynamic strategy above). Below, we describe how each of these three components are represented, one by one, and give examples of how some trading constraints can be modeled.

The difference between the static and the dynamic parts requires different modeling strategies. For the static strategies, the trader simply needs to decide an allocation or a strike price. This can be modelled with a set of trainable parameters, which does not take an input, \textit{i.e.,} these allocations are not functions of some input but instead constant vectors. For the dynamic trading strategy, which, at each trading date, is a function of the current wealth, we use a deep neural network taking the current wealth as input and outputs the stock allocation. The set of admissible trading strategies $\Pi=\{\Pi^\beta,\Pi^K,\Pi^{\alpha_0},\Pi^{\alpha_1},\ldots,\Pi^{\alpha_{N-1}}\}$, where we note that $\Pi^{\alpha_1},\ldots,\Pi^{\alpha_{N-1}}$, may also depend on the stock process. For instance, there may be a constraint on how high the exposure can be in a single stock which also depends on the evolution of that specific stock value.
\newline\newline
\textbf{Static option strategy} ($\beta$):\newline
The number of trainable parameters needed to represent $\beta$ is the same as the number of options, \textit{i.e.,} $N^\text{options}$, implying that $\theta^\beta\in\R^{N^\text{options}}$. Moreover, an activation function $\boldsymbol{a}^\beta\colon\R^{N^\text{options}}\to\Pi^\beta$ is applied to ensure that the trading constraints are satisfied. The obtained $\beta-$strategy is then given by
\begin{equation*}
    \hat{\beta}=\boldsymbol{a}^\beta(\theta^\beta).
\end{equation*}
For instance, if there are no constraints on how we are allowed to trade in the options, then $\Pi^\beta=\R^{N^\text{options}}$ and $\boldsymbol{a}^\beta$ would be the componentwise identity function. 
A more realistic constraint would be to not allow short selling and a maximum total amount of $\beta^\text{max}$ plus a maximum amount of $\frac{\beta^\text{max}}{N^\text{options}}$ in each option, which would give $\Pi^\beta=\big[\boldsymbol{0}^{N^\text{options}},\boldsymbol{1}^{N^\text{options}}\times\frac{\beta^\text{max}}{N^\text{options}}\big]$, where $\boldsymbol{0}^{N^\text{options}}$ and $\boldsymbol{1}^{N^\text{options}}$ are vectors of zeros and ones, respectively. A natural way to achieve this is to set \begin{equation}\label{opt_net}
    \boldsymbol{a}^\beta(\theta^\beta)=\frac{\beta^\text{max}}{N^\text{options}}\times\text{sigmoid}(\theta^\beta),
\end{equation}
where $\text{sigmoid}(\cdot)$ is the componentwise sigmoid function, which in each component is given by $(1-\text{e}^{-x})^{-1}$.
\newline\newline
\textbf{Static strike price strategy} ($K$):\newline
Similar to $\beta$ above, the number of trainable parameters for the strike prices is $N^\text{options}$ and we use the notation $\boldsymbol{a}^K\colon\R^{N^\text{options}}\to\Pi^K$ with the $K$-strategy
\begin{equation*}
    \hat{K}=\boldsymbol{a}^K(\theta^K).
\end{equation*}
In practice, there is a discrete set of strike prices available in the market, and a reasonable model is to consider each option strike price in some interval $[K^\text{low},K^\text{high}]$, implying that $\Pi^K=[K^\text{low}\times\boldsymbol{1}^{N^\text{options}},K^\text{high}\times\boldsymbol{1}^{N^\text{options}}]$. A natural choice for such an activation function is given by\begin{equation}\label{K_net}
    \boldsymbol{a}^K(\theta^K) = (K^\text{high}-K^\text{low})\odot\text{sigmoid}(\theta^K) \oplus K^\text{low},
\end{equation}  
where $\odot$ and $\oplus$ denote componentwise multiplication and addition, respectively.
\newline\newline
\textbf{Dynamic stock and bond strategy} ($\alpha$):\newline
Since the allocation into the bond and the stocks changes at each trading date, we need to approximate the sequence $\alpha=\{\alpha_0,\ldots,\alpha_{N-1}\}$. At $t_0$, the allocation is deterministic, since we assume a fixed initial wealth, therefore $\alpha_0$ needs to be treated differently from $\alpha_n$, for $n\in\{1,\ldots, N-1\}$. The number of trainable parameters in the representation of $\alpha_0$ is $N^\text{stocks}$ (the amount traded in the bond is given to maintain the self-financing constraint) and we use the notation $\boldsymbol{a}^{\alpha_0}\colon\R^{N^\text{stocks}}\to\Pi^{\alpha_0}$, with $\alpha_0-$strategy
\begin{equation*}(\hat{\alpha}_0^1,\ldots,\hat{\alpha}_0^{N^\text{stocks}})^\top=\boldsymbol{a}^{\alpha_0}(\theta^{\alpha_0}).
\end{equation*}
A typical setup at $t_0$ can be that the trader is allowed to allocate a specific amount within a specified interval $[A_0^\text{low},A_0^\text{high}]$, which gives $\Pi^{\alpha_0}=[A_0^\text{low}\times\boldsymbol{1}^{N^\text{stocks}},A_0^\text{high}\times\boldsymbol{1}^{N^\text{stocks}}]$. Note that $\alpha_0^k$ is the amount of stock $k$ in the portfolio and not the amount of cash allocated into stock $k$. Therefore, it is $A_0^k$ that is supposed to be within $[A_0^\text{low},A_0^\text{high}]$. On the other hand, since we have assumed that the initial values of all assets equals 1, the definition still holds. Similar to above, we can then use the activation function 
\begin{equation}\label{act_stock_alo_0}
    \boldsymbol{a}^{\alpha_0}(\theta^{\alpha_0}) = (A_0^{\text{high}}-A_0^\text{low})\odot\text{sigmoid}(\theta^{\alpha_0}) \oplus A_0^{\text{low}}.
\end{equation}  
Finally, $\hat{\alpha}_0$ (which also include the initial bond allocation) is given by \begin{equation*}
    \hat{\alpha}_0=\big(\hat{\alpha}^0_0,\,\boldsymbol{a}^{\alpha_0}(\theta^{\alpha_0}) ^\top\big)^\top,
\end{equation*}
where $\hat{\alpha}_0^0=\hat{x}_0-\sum_{k=1}^{N^\text{stocks}}\hat{\alpha}_0^k$, with $\hat{x}_0=x_0^\text{IC}-\hat{y}_0$ and $\hat{y}_0=\sum_{k=1}^{N^\text{options}}\hat{\beta}^k$.

The above trading strategies all have in common that they are decided at time $t_0$ before any randomness enters the system, and hence, they do not depend on previous performance. For trading after $t_0$, on the other hand, we also need to take previous performance, as well as random realization of the asset processes, into account. Therefore, we use fully connected neural networks of the form from Section \ref{FCNN} \begin{equation*}
(\hat{\alpha}_n^1,\ldots,\hat{\alpha}_n^{N^\text{stocks}})^\top=\phi(S_{t_n}\,;\theta^{\alpha_n}).
\end{equation*}
Given a parametrization $\theta^{\alpha_n}$, the above is a mapping from $S_{t_n}$, which represents the value of all the stocks at $t_n$, to the new stock allocation, or $\alpha_n-$strategy. On the other hand, given $S_{t_n}$, the neural network is a mapping from the parameter space to the admissible strategy space, \textit{i.e.,} $\phi(S_{t_n};\,\cdot\,)\colon\R^{\mathfrak{D}^{\theta_n}}\to\Pi^{\alpha_n}$, where $\mathfrak{D}^{\theta_n}$ is the number of trainable parameters, as specified in Section \ref{FCNN}. The activation functions in the interior layers can be chosen arbitrarily, but as for the strategies above, the activation function in the output layer should be chosen such that the trading constraints are satisfied. With the same trading constraints as for $t_0$, we have $\Pi^{\alpha_n}=[A_n^\text{low}\times\boldsymbol{1}^{N^\text{stocks}}\varobslash S_{t_n},A_n^\text{high}\times\boldsymbol{1}^{N^\text{stocks}}\varobslash S_{t_n}]$, where $\varobslash$ denotes componentwise division. The reason for dividing by the stock process above is that it is usually not the amount of each stock that matters from a constraint perspective, but rather the value of the traders position in each stock. Below, we specify a possible activation function in the output layer, which guarantees that the trading constraints are obeyed \eqref{act_stock_alo_0}.
\begin{equation}\label{act_stock_alo_n}
    \boldsymbol{a}^{\alpha_n}(\chi) = \big[(A_n^{\text{high}}-A_n^{\text{low}})\odot\text{sigmoid}(\chi) \oplus A_n^{\text{low}}\big]\varobslash S_{t_n},
\end{equation}
where $\chi=w_\mathfrak{L}^\top L_{\mathfrak{L}-1}+b_\mathfrak{L}$, with $L_{\mathfrak{L}-1}$ is the vector from layer $\mathfrak{L}-1$ (with slight abuse of notation) and $w_\mathfrak{L}$ and $b_\mathfrak{L}$ are the weight and bias vectors, respectively. Similar to above, we have $\hat{\alpha}^n$, which is given by \begin{equation*}
\hat{\alpha}_n=\big(\hat{\alpha}^n_0,\phi(S_{t_n};\theta^{\alpha_n})^\top\big)^\top,
\end{equation*}
where $\hat{\alpha}_n^0=(B_{t_n})^{-1} \big(\hat{x}_{t_n}-\sum_{k=1}^{N^\text{stocks}}\hat{\alpha}_n^kS_{t_n}^k\big)$, with $\hat{x}_{t_n}=\hat{\alpha}_n^0B_n+\sum_{k=1}^{N^\text{stocks}}\hat{\alpha}_{n-1}^kS_{t_n}^k$.

\subsubsection{Optimization problem with neural networks}
Summarizing all the trainable parameters from above, we conclude that our neural network based trading strategy $(\hat{\beta},\hat{K},\hat{\alpha})$ is determined by the parameters $\theta = (\theta^\beta,\theta^K,\theta^{\alpha_0},\theta^{\alpha_1}\ldots,\theta^{\alpha_{N-1}})$, where each component is vectorized before concatenation. Parameters $\theta$ as given above contain all the trainable parameters for a sequence of networks which together control the trading strategy. We denote by $\Theta$ the set of all neural network parameters such that $\theta\in\Theta$ produces an admissible trading strategy. 

Below, a list of the strategy evaluated at one random realization of the underlying asset process is given by\\
$\big\{\boldsymbol{a}^K(\theta^K), \boldsymbol{a}^\beta(\theta^\beta),\boldsymbol{a}^{\alpha_0}(\theta^{\alpha_0}), \phi_1(S_{t_1};\theta^{\alpha_1}),\ldots,\phi_{N-1}(S_{t_{N-1}};\theta^{\alpha_{N-1}})\big\}$,
and the full neural network based optimization problem is defined by
\begin{equation}\label{nn_opt_prob}
\begin{dcases}
\underset{\theta\in\Theta}{\mathrm{maximize}}
 = \bar{U}^M(\theta),\quad \text{where the } M \textit{ i.i.d. }\text{random variables follow},\\
R(S\,;\theta) = R_{\text{SB}}(S\,;\theta) + R_\text{O}(S\,;\theta)\\
R_{\text{SB}}(S\,;\theta)= \hat{x}_{t_N} - \hat{x}_0 - \sum_{k=1}^{N^\text{stocks}}\text{TC}^k,\quad R_\text{O}(S\,;\theta)=\hat{y}_{t_N}-\hat{y}_0,\\
\hat{x}_{t_N} = \hat{x}_0 + \sum_{n=0}^{N}\I_{\{x>0\}}\big(\hat{x}_{t_n}\big)\big[\hat{\alpha}_n^0(B_{t_{n+1}}-B_{t_n}) + \sum_{k=1}^{N^\text{stocks}}\hat{\alpha}_n^k(S_{t_{n+1}}^k - S_{t_n}^k)\big],\quad \hat{x}_0=\hat{x}_0^\text{IC}-\hat{y}_0,\\
\hat{y}_{t_N} = \sum_{i=1}^{N^\text{options}}\hat{\beta}^iV^i(T,S_{t_N}\,;\,\hat{K}^i),\quad \hat{y}_0 = \sum_{i=1}^{N^\text{options}}\hat{\beta}^iV^i(0,S_0\,;\,\hat{K}^i),\\
\hat{\alpha}_0^0=\hat{x}_0-\sum_{k=1}^{N^\text{stocks}}\hat{\alpha}_0^k,\quad (\hat{\alpha}_0^1,\ldots,\hat{\alpha}_0^{N^\text{stocks}})^\top=\boldsymbol{a}^{\alpha_0}(\theta^{\alpha_0}),\quad \hat{\beta}=\boldsymbol{a}^\beta(\theta^\beta),\quad \hat{K}=\boldsymbol{a}^K(\theta^K),\\
\hat{\alpha}_n^0=\frac{1}{B_{t_n}}\big(\hat{x}_{t_n}-\sum_{k=1}^{N^\text{stocks}}\hat{\alpha}_n^kS_{t_n}^k\big),\quad (\hat{\alpha}_n^1,\ldots,\hat{\alpha}_n^{N^\text{stocks}})^\top=\phi(S_{t_n}\,;\theta^{\alpha_n}).
\end{dcases}
\end{equation}
In the above, $\bar{U}^M(\theta)=u(\bar{R}^M(\theta))$, where $\bar{R}^M(\theta)=\{R(S(1);\theta),\ldots,R(S(M);\theta)\}$ is the empirical distribution from $M$ samples distributed as described in \eqref{nn_opt_prob} above.

\subsubsection{Pseudo-code}\label{pseudo_code}
Denote the number of training samples, batch size and the number of epochs by $M_\text{train}$, $M_\text{batch}$, $M_\text{epoch}\in\N_+$, respectively. Pseudo-code aiming to describe the proposed algorithm is given below. For notational convenience, the pseudo-code describes the algorithm in the special case when $M_\text{epoch}=1$, but the extension to $M_\text{epoch}>1$ can simply be achieved by repeating the procedure $M_\text{epoch}$ times.
\begin{algorithm}[htp]\label{algorithm}
	\KwIn{Initialization of neural network parameters, $\{\theta^\beta(1), \theta^K(1),\theta_0^\alpha(1),\ldots,\theta_{N-1}^\alpha(1)\}$, initial state $S_0$, initial cash $\hat{x}^\text{IC}$, and for $0\leq m\leq M_\text{train}$ and $1\leq n\leq N-1$, asset state $S_{t_n}(m)$.}
	\KwOut{Approximation of the trading strategy, determined by $\beta, K$ and $(\alpha_t)_{t\in \T^N}$.}
  \For{$k=1,2,\ldots,K_\mathrm{batch}$, where $K_\mathrm{batch}=M_\mathrm{train}/M_\mathrm{batch}$ \emph{is the number of batches.} \emph{(should be carried out sequentially)}}{
\For{ $m\in\{1,\ldots,M_\mathrm{batch}\}$ \emph{(may be carried out in parallel)}}{
	\vspace{0.2cm}
	$\hat{\beta}=\boldsymbol{a}^\beta(\theta^\beta(k)),\quad \hat{K}=\boldsymbol{a}^K(\theta^K(k)),\quad\hat{y}_0 = \sum_{i=1}^{N^\text{options}}\hat{\beta}^iV^i(0,S_0\,;\,\hat{K}^i),\quad\hat{x}_0=\hat{x}_0^\text{IC}-\hat{y}_0,\quad(\hat{\alpha}_0^1,\ldots,\hat{\alpha}_0^{N^\text{stocks}})^\top=\boldsymbol{a}^{\alpha_0}(\theta^{\alpha_0}(k)),\quad\hat{\alpha}_0^0=\hat{x}_0-\sum_{k=1}^{N^\text{stocks}}\hat{\alpha}_0^k,$ \newline
 $\hat{x}_{t_1}(m) = \hat{x}_{t_0} + \I_{\{x>0\}}\big(\hat{x}_{t_0}\big)\big[\hat{\alpha}_0^0(B_{t_{1}}-B_{t_0}) + \hat{\alpha}_0^k\big(S_{t_{1}}(m) - S_0\big)\big]$\\[1.5ex]
	\For{$n=1, \ldots, N-1$ \emph{(should be carried out sequentially)}}{
 $ (\hat{\alpha}_n^1(m),\ldots,\hat{\alpha}_n^{N^\text{stocks}})^\top=\phi(S_{t_n}(m)\,;\theta^{\alpha_n}(k)),\quad \hat{\alpha}_n^0(m)=\frac{1}{B_{t_n}}\big(\hat{x}_{t_n}-\sum_{k=1}^{N^\text{stocks}}\hat{\alpha}_n^kS_{t_n}^k(m)\big),$
	$\hat{x}_{t_{n+1}}(m) = \hat{x}_{t_n}(m) + \I_{\{x>0\}}\big(\hat{x}_{t_n}(m)\big)\big[\hat{\alpha}_n^0(m)(B_{t_{n+1}}-B_{t_n}) + \hat{\alpha}_n^k(m)\big(S_{t_{n+1}}(m) - S_{t_n}(m)\big)\big]$
	}
  $R(S(m)\,;\theta(k)) = R_{\text{SB}}(S(m)\,;\theta(k)) + R_\text{O}(S(m)\,;\theta(k))$\\$
R_{\text{SB}}(S(m)\,;\theta(k))= \hat{x}_{t_N}(m) - \hat{x}_0 - \sum_{k=1}^{N^\text{stocks}}\text{TC}^k(m),\quad R_\text{O}(S(m)\,;\theta(k))=\hat{y}_{t_N}(m)-\hat{y}_0,$
 }    \vspace{0.2cm}
 $\theta(k)=\{\theta^\beta(k),\theta_K(k),\theta_{\alpha_0}(k),\theta_{\alpha_1}(k)\ldots,\theta^{\alpha_{N-1}}(k)\},\  \text{(trainable parameters)}$  \\
 $\bar{R}^{M_\text{batch}}(\theta(k))=\{R(S(1)\,;\theta(k)),\ldots, R(S(M_\text{batch})\,;\theta(k))\}$ (empirical distribution).
  \\ \vspace{0.2cm}
	$\bar{U}^{M_\text{batch}}(\theta(k))= u\big(\bar{R}^{M_\text{batch}}(\theta(k))\big)$ (Loss-function)\\
	$\theta(k+1)\leftarrow\argmin_{\theta}u(\bar{R}^{M_\text{batch}}(\theta))$ (some optimization algorithm, usually of gradient ascent type).
 }
\caption{Pseudo-code of one epoch of the neural network training.}
\label{alg:osm}
\end{algorithm}

\section{Numerical experiments}\label{num_exp}
In this section, we aim to demonstrate the effectiveness and flexibility of the proposed algorithm. In previous sections, the asset process has not been fully specified and the reason for this is that any process, computer generated or real world data, can be used. In this section, on the other hand, we need to decide how to generate the training data and we use the jump-diffusion SDE. Moreover, the fully implementable model includes the market frictions as well as the European call and put options, which are introduced below. A special case, which fits into this framework, is the mean-variance (MV) portfolio optimization problem, which admits to a closed-form analytic solution (an admissible trading strategy in closed form, which is optimal in the sense of the objective function). This solution is used here as a reference for our approximate trading strategy for the MV portfolio optimization problem. In subsequent sections, we go beyond classical MV problems and consider market frictions, other objective functions as well as more general asset dynamics. This leads to interesting investment strategies in which combinations of a bond, stocks and options seem optimal. 

The asset model used in this paper is a Geometric Brownian motion with multiplicative jumps, which is given by
\begin{numcases}{}\label{S_t}
    \d S_t = b\odot S_t\d t+ \sigma S_t\odot\d W_t + J\odot S_t\odot\d X_t;\quad S_0=(1,\ldots,1)^\top,\\\label{B_t}
    \d B_t = rB_t\d t; \quad B_0=1.
\end{numcases}
Here, $T\in[0,\infty)$, $N^\text{stocks}\in\N$, $(W_t)_{t\in[0,T]}$ is an $N^\text{stocks}$-dimensional standard Brownian motion, $J$ is an $N^\text{stocks}$-dimensional multivariate normally distributed random variable with mean $\mu_J\in\R^{N^\text{stocks}}$ and covariance $\Sigma_J\in\R^{N^\text{stocks}\times N^\text{stocks}}$, $(X_t)_{t\in[0,T]}$ is an $N^\text{stocks}$-dimensional Poisson process parametrized by $\xi_p\in\R_+^{N^\text{stocks}}$ and model parameters $b\in\R^{N^\text{stocks}}$ and $\sigma\in\R^{N^\text{stocks}\times N^\text{stocks}}$. The initial values $S_0$ and $B_0$ are scaled to unity for the purpose of simplicity. 

Setting $x_0=1$ and using a binary parameter for the no-bankrupcy-constraint $\text{NB}\in\{0,1\}$, we have
\begin{align}\nonumber
    \d x_t=& \big(1-\text{NB}\times\I_{\{x\leq 0\}}(x_t)\big)\Big[\big(r\alpha_t^0B_t + \sum_{i=1}^{N^\text{stocks}}b_i\alpha_t^iS_t^i\big)\d t \\
    &+ \sum_{i=1}^{N^\text{stocks}}\sum_{j=1}^{N^\text{stocks}}\sigma_{ij}\alpha_t^iS_t^i\d W_t^j + \sum_{i=1}^{N^\text{stocks}}\alpha_t^iS_t^iJ_t^i\d X_t^i\Big]\\ \label{x_t}
    \nonumber=&\big(1-\text{NB}\times\I_{\{x\leq0\}}(x_t)\big)\Big[\big(rx_t + \sum_{i=1}^{N^\text{stocks}}(b_i - r)N^i_t\big)\d t \\
    &+ \sum_{i=1}^{N^\text{stocks}}\sum_{j=1}^{N^\text{stocks}}\sigma_{ij}A_t^i\d W_t^j + \sum_{i=1}^{N^\text{stocks}}A_t^iJ_t^i\d X_t^i\Big],
\end{align}
where we recall that $A_t^i = \alpha_t^iS_t^i$. The problem here is independent of the the asset process $S$. This allows for a simplification of the feedback map $S_n\mapsto\hat{\alpha}_n$, which may be reduced to $\hat{x}_n\mapsto\hat{\alpha}_n$. In practice, this implies that the neural networks can map a one-dimensional process $\hat{x}_n$ to the stock allocation instead of the $N^\text{stocks}$ asset process. It has been verified that these two neural network structures produce close to identical results with only a minor improvement in the computational time for the latter. This implies that it is reasonable to stick to the more general form, which is presented in this paper. Transaction costs are expressed according to \eqref{TC}. Note that the above formulation is expressed in a time-continuous fashion, while in Section \ref{sec2}, the allocation process as well as $x$, are expressed in a a time discrete way. The reason for this is that the classical MV-optimization problem admits to a closed-form solution for the allocation process which offers the opportunity to compare with a reference solution in this special case. Moreover, by considering a piecewise continuous allocation process, the formulations coincide (up to a discrete approximation of the geometric Brownian motion with jumps).
\newline\newline
\noindent \textbf{General neural network settings}
Although any function approximator can be used in \eqref{nn_opt_prob}, we choose a sequence of neural networks (as described in the pseudo-code in Algorithm \ref{alg:osm}). In this paper, the neural networks are seen merely as tools to solve the problem. Therefore, the focus is not placed on optimizing the structure or parameter choices but rather on finding a setting which is not too much problem dependent. We therefore use the same settings, which are given below, for all problems considered in this paper. Following the notation in Sections \ref{FCNN} and \ref{pseudo_code}, the total number of training samples is set to $M_\text{train}=2^{22}$, the batch size to $M_\text{batch}=2^{12}$, the number of epochs to $M_\text{epoch}=10$ and the number of layers to $\mathfrak{L}=4$. For the interior layers, \textit{i.e.,} $\ell\in\{2,3\}$, we set the number of nodes to $\mathfrak{N}_\ell=20$ and the activation functions $\mathbf{a}_\ell(\cdot)=\text{ReLU}(\cdot)$. The input dimension $\mathfrak{D}_\text{input}=1$ and the output dimension $\mathfrak{D}_\text{output}$, as well as the activation function in the output layer depend on the trading constraints and are specified for each specific problem below. We use 0.01 as initial learning rate and after two batches it decreases with a factor $\text{exp}(-0.5)$ for each new batch. We do not apply regularization or normalization techniques.

\subsection{Classical continuous mean-variance optimization}\label{NR_MV}
As a first example, we consider the classical mean-variance optimization problem. This means that the asset process \eqref{S_t} is a pure geometric Brownian motion. This can fit our framework, for instance, by setting the $\mu_J$
and $\Sigma_J$ to a zero vector and a zero matrix, respectively. Moreover, trading should be carried out without transaction costs and no-bankruptcy constraint, \textit{i.e.,} setting the courtage, $C=0$ and $\text{NB}=0$ and there should be no constraints for short selling, leverage or bankruptcy. Here trading in the options is not allowed and the algorithm used is D-TIP. The MV objective function is given by \begin{equation*}
    U(\pi)=E[x_T] - \lambda\text{Var}[x_T],
\end{equation*}
where $\lambda>0$ controls the risk aversion.
From \cite{DMVO_LQf}, we know that there exists an analytic solution to this problem in terms of a closed-form expression for the optimal allocation as well as a theoretical optimal mean and variance of the terminal wealth, \textit{i.e.,} an infimum of the objective function.

In the example below, the following parameter values are used \begin{equation}\label{parameter_values}
   T=2, N=20, r=0.06, b=\begin{pmatrix}0.08\\ 0.07\\ 0.06\\ 0.05\\0.04\end{pmatrix}, \sigma = \begin{pmatrix}
    0.23&0.05&-0.05&0.05&0.05\\
        0.05&0.215&0.05&0.05&0.05\\
    -0.05&0.05&0.2&0.05&0.05\\
    0.05&0.05&0.05&0.185&0.05\\
    0.05&0.05&0.05&0.05&0.17\\
\end{pmatrix}, \lambda = 1.104.
\end{equation}
The parameter setting as given above leads to a non-symmetric problem with asset dynamic with correlated components, \textit{i.e.,} each stock is correlated to the other stocks. Moreover, we vary between positive correlation and negative correlation, which typically is the case in factor investing \cite{factor_investing}.
\newline\newline
\noindent \textbf{Problem specific neural network settings}
In addition to the general neural network settings from Section \ref{NN_spec} and the introduction of Section \ref{num_exp}, we here give the finer, problem specific, details. Since D-TIP does not allow options in the portfolio, we only consider the networks that compute the dynamic stock and bond strategy. 

We have one neural network per allocation point and the requirements are \textit{i}) the output dimension should coincide with the number of stocks, $N^\text{stocks}$, and \textit{ii}) since we do not apply any trading constraints, the range for each individual stock allocation is $\R$ and hence the activation function in the output layer should have $\R$ as its range for each component. This is achieved
by adjusting \eqref{act_stock_alo_0} and \eqref{act_stock_alo_n} to \begin{align*}
    \boldsymbol{a}^{\alpha_0}(\theta^{\alpha_0})&=\theta^{\alpha_0},\quad \text{with } \theta^{\alpha_0}\in\R^{N^\text{stocks}},\\
    \boldsymbol{a}^{\alpha_n}(\chi) &=\chi, \quad \text{with } \chi\in\R^{N^\text{stocks}},\quad\text{for }n\in\{1,2,\ldots,N-1\},
\end{align*}
where $\chi$ is the output of layer $3$, \textit{i.e.,} the output of last hidden layer in our neural networks.

\begin{figure}[htp]
\centering
\begin{tabular}{ccc}
          \includegraphics[width=80mm]{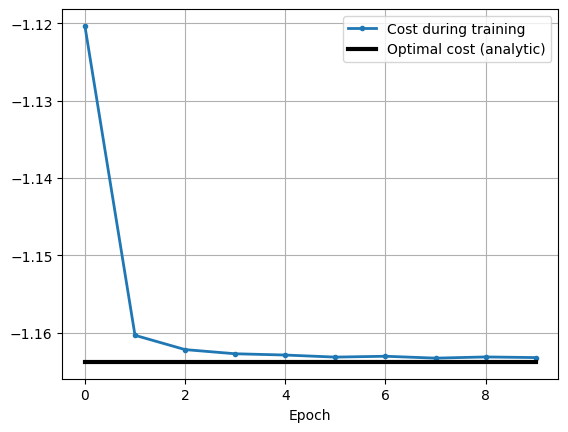}&  \includegraphics[width=80mm]{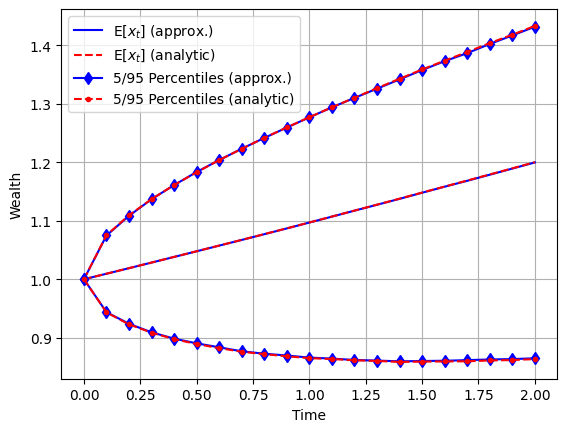}\\ \includegraphics[width=80mm]{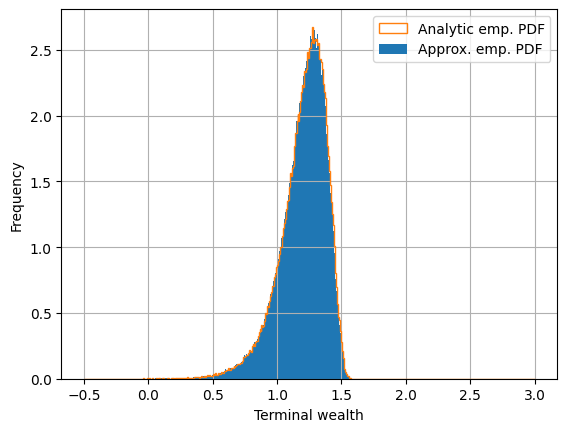}&
          \includegraphics[width=80mm]{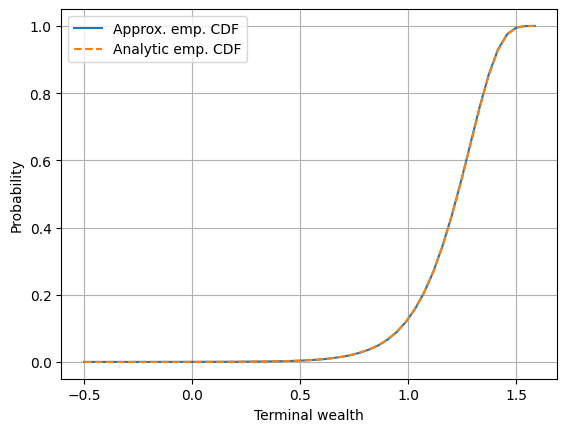}
\end{tabular}
\caption{\textbf{Upper left:} Convergence of the loss to the analytic counterpart with respect to the number of training epochs. \textbf{Upper right:} Comparison of our approximation and the reference solution. Upper lines, middle lines and lower lines represent the 95th percentiles, the mean values and the 5th percentiles, respectively. \textbf{Lower left:} Comparison of the empirical pdfs of our approximation and the reference solution. \textbf{Lower right:} Comparison of the empirical CDFs of our approximation and the reference solution.}\label{MV_reference}
\end{figure}

The optimal value of the objective function is approximately $1.1637$ and in Figure \ref{MV_reference}, top left, we see the convergence of our approximation to the true value with respect to the number of epochs\footnote{Since our data is synthetic we could have generated new data instead of reusing data in each epoch. On the other hand, since our algorithm is data driven, it would be possible to use real world data which makes it desirable to be able to reuse data.}. It should be emphasized that the implementation results in a discrete approximation of a continuous problem. By increasing $N$ the approximation should converge with respect to the step size $h=T/N$, but to investigate this is beyond the scope of this paper. The top right plot shows the expected value as well as the 5:th and 95:th percentiles of the wealth as a function of time. We see that the approximation corresponds well to the analytic counterpart. The plots at the bottom display a comparison of the empirical pdfs and CDFs, respectively, in which it is clear that our approximation corresponds well to the empirical distributions obtained from the analytic allocation strategy.

\subsection{Beyond MV, with market frictions and jumps}
In this subsection, we consider the full generality offered by the asset model from \eqref{S_t}-\eqref{B_t}, as well as transaction costs, no bankruptcy constraint and we allow for trading in European call and put options. When also options are allowed to trade in, the algorithm used is D-TIPO. The parameter values from \eqref{parameter_values} are reused and we set $\lambda_J=0.05$, $\mu_J=(0,\ldots,0)^\top$, $\Sigma_J=\text{diag}(0.2,\ldots,0.2)$, $\text{NB}=1$ and $C=0.005$. This means that we have an asset price process with symmetric jumps around zero for each individual asset and a courtage of $0.5\%$ for each trade. This could seem high, but an alternative interpretation of $C$ is as a penalizing term for too heavy reallocation (which is something that for instance pension funds want to avoid). With a slight abuse of notation, we denote the discretized version of $\eqref{S_t}$ by $S$ but keep in mind that we must rely on an approximation in this section. In all our examples, we use $N$ (the number of allocation dates) as time steps in an Euler--Maruyama scheme, however, if necessary, a finer mesh can be used. In turn, this also implies that \eqref{x_t} has to be approximated and this is done with the Euler--Maruyama scheme as well. When there is no risk of confusion, we simply denote the discretized versions of $S$ and $x$ by \begin{equation*}
    S_n = S_{t_n},\quad x_n=x_{t_n},
\end{equation*}
where $t_n$ is the $n$:th allocation date.

For each stock, $S^i$, $i\in\{1,\ldots,N^\text{stocks}\}$, we allow trading in one European call option and one European put option with terminal time $T$ and strike $K_i$. For $i\in\{1,\ldots,N^\text{stocks}\}$, denoted by $\bar{V}^i(t,S_t^i;K^i)$, the risk neutral price of a call option, and for $i\in\{\ldots,N^\text{stocks}+1,\ldots,2\ldots,N^\text{stocks}\}$, we denote by $\bar{V}^{N^\text{stocks}+i}(t,S_t^i;K^{N^\text{stocks}+i})$ a put option with $S^i$ as underlying. Note that a drift correction term needs to be added to the asset dynamics to obtain a risk neutral measure, which is necessary in order to make the option values martingales under the jump-diffusion SDE, see \textit{e.g.,} \cite{JD-SDE}. We then have that $\bar{V}^i(T,S_T^i;K^i)=\text{max}(0,S_T^i-K^i)$ and $\bar{V}^{N^\text{stocks}+i}(T,S_T^i;K^{N^\text{stocks}+i})=\text{max}(0,K^{N^\text{stocks}+i}-S_T^i)$. We want to work with normalized entities and define \begin{align*}
    V^i(0,S_0^i;K^i)&=1, \quad V^i(T,S_T^i;K^i)= \frac{\text{max}(0,S_T^i - K^i)}{\bar{V}^i(0,S_0^i;K^i)},\\
        V^{N^\text{stocks}+i}(0,S_0^i;K^{N^\text{stocks}+i})&=1, \quad V^{N^\text{stocks}+i}(T,S_T^i;K^{N^\text{stocks}+i})=\frac{\text{max}(0,K^{N^\text{stocks}+i}-S_T^i)}{\bar{V}^{N^\text{stocks}+i}(0,S_0^i;K^{N^\text{stocks}+i})}.
\end{align*}
In this section, we use the objective function from \eqref{MV_ES_ES_objective}. The main reason is that this extended version of the mean-variance objective function allows us to better control the tails of the final wealth distribution (or in fact, of the return $R$). We want to use an objective function which has similarities to mean-variance since it is desirable to achieve a high mean and a low variance of the terminal wealth. On the other hand, we want to be able to better control the tail distribution. With this in mind, we add two terms aiming to increase the expected value in the tails. Below we give the theoretical version of the used objective function, but when implemented, we always resort to $\bar{U}^M(\pi)$, which is the $M$-sample empirical counterpart to $U(\pi)$.
\begin{equation*}
    U(\pi) = \E[R] - \lambda_1\text{Var}[R] + \lambda_2\text{ES}^-_{ p_1}(R) + \lambda_3\text{ES}^+_{ p_2}(R).
\end{equation*}
We use $p_1=0.01$ which means that we penalize low values of the expected return of the worst 1\% performance of the portfolio. For the upper tail, we want to maximize the expected return of the 5\% best outcomes and therefore set $ p_2=0.95$. The weights are set to $\lambda_1=0.552$, $\lambda_2=0.276$ and $\lambda_3 = 0.110$. 
\newline\newline
\noindent \textbf{Problem specific neural network settings}
In addition to the general neural network settings from Section \ref{NN_spec} and the introduction of Section \ref{num_exp}, we here give the finer details which are problem specific. The D-TIPO algorithm does not only return a dynamic trading strategy for the bond and the stocks, as in classical MV-optimization, but also static strategies for allocations into a set of options as well as a strike price for each option. For the option allocation, we follow \eqref{opt_net} and set $\beta_\text{max}=1$. This implies that the maximum total amount of allocation into the options is 100\% of the initial wealth. One problem with \eqref{opt_net} is that when the number of stocks increases, the possible allocation into each option decreases. Since $N_\text{options}=10$ in the example considered in this section, the maximum amount allocated into each specific option is $0.1$. We therefore use a slight modification of \eqref{opt_net}, given by
\begin{equation*}
    \boldsymbol{a}^\beta(\theta^\beta)=\frac{1}{\text{max}\{1,\hat{\beta}\}}\times\hat{\boldsymbol{a}}^\beta(\theta^\beta),
\end{equation*}
where $    \hat{\boldsymbol{a}}^\beta(\theta^\beta)=\frac{\beta^\text{max}}{N^\text{options}}\times\text{sigmoid}(\theta^\beta)$ and $ \hat{\beta}=\sum_{i=1}^{N_\text{options}}[ \hat{\boldsymbol{a}}^\beta(\theta^\beta)]_i.$ In this way, the allocation range into each option is $[0,\beta^\text{max}]$ while keeping the total allocation range (the sum of the allocations into each option) to $[0,\beta^\text{max}]$ as well. 

For the strike prices, we follow \eqref{K_net} and use $K^\text{low}=(0.75,0.75\ldots,0.75)^\top$ and $K^\text{high}=(1.25,1.25\ldots,1.25)^\top$, \textit{i.e.,} setting the range for strike prices to between $75\%$ to $125\%$ of the stock price at the initial time.

Finally, for $n\in\{0,1,\ldots,N-1\}$, \eqref{opt_net} is used and we set $\alpha_n^\text{low}=-2x_n$ and $\alpha_n^\text{high}=2x_n$ implying that we can allocate into each stock between $-200\%$ and $200\%$ of the total value of the stocks and the bond. 

\subsubsection{Evaluation of the results}
In this section, we aim to show that for a more sophisticated objective function, it can be beneficial to include options to a portfolio of stocks and a bond.

\begin{figure}[htp]
\centering
\begin{tabular}{ccc}
          \includegraphics[width=80mm]{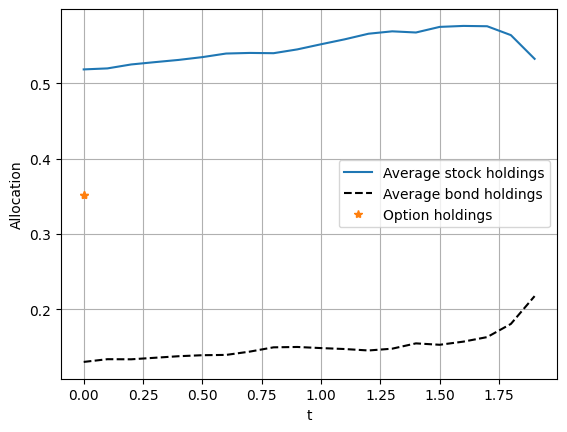}&  
          \includegraphics[width=80mm]{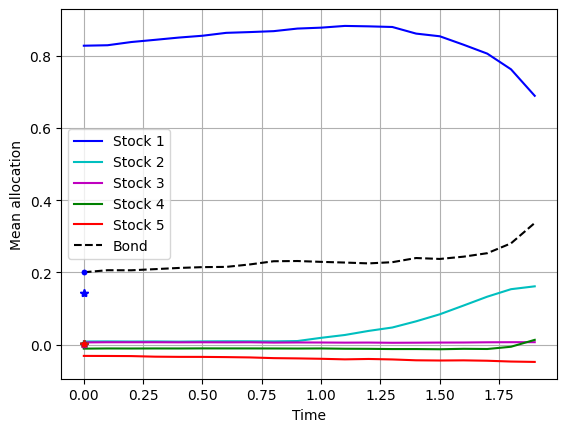}
\end{tabular}
\caption{\textbf{Left:} Average allocation to stocks, bond and options over time. \textbf{Right:} Average allocation to stocks, bond and options over time for each stock. Asterisks and bullets represent call and put option holdings, respectively.}\label{options}
\end{figure}

\begin{figure}[htp]
\centering
\begin{tabular}{c}
          \includegraphics[width=100mm]{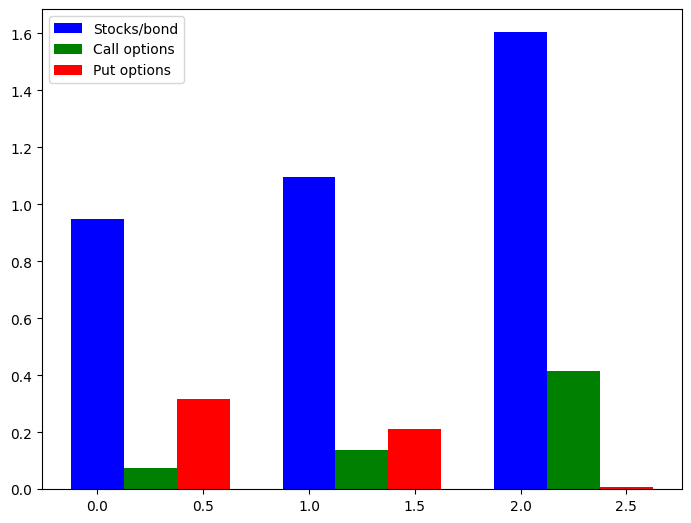}
          \end{tabular}
\caption{ Contribution to the portfolios for terminal wealth less than 1.03 (33\% of the outcomes), between 1.03 and 1.12 (41\%), and above 1.12 (26\%).}\label{stocks_opt}
\end{figure}

To begin with, we display the kind of strategy D-TIPO generates. In Figure \ref{options} to the left, the average total allocation into the stocks (the sum of all stock positions) and the bond over time as well as the total allocation to the options (the sum of all option positions) is displayed. We see that approximately 35\% of the initial wealth is allocated to the options and the remainder to the stocks and the bond. In Figure \ref{options} to the right, the average allocations are displayed for each individual stock and we note that the main allocation is into stock 1, for the stock itself as well as for the call and put options.
The purpose of Figure \ref{stocks_opt} is to gain insight into differences between portfolios with different levels of performance, \textit{e.g.,} is there a difference in the contribution from the different asset classes (stocks, bonds and options) for the best contra worst performing portfolios? In Figure \ref{stocks_opt} the average contributions of the stocks and the bond, the call options and put options are displayed for portfolios in three different performance ranges. From left to right, we see portfolios with terminal wealth less than 1.03, between 1.03 and 1.12, and above 1.12, respectively. The strike prices are optimized by the neural networks to 0.75 for all call options and 1.25 for all put options, \textit{i.e.,} deep in the money. This enhances the viewpoint that the main purpose of the options in the portfolio is to cover up for tail events of the stocks rather than to leverage potential upside. We also note that for the best performing outcomes, the main option contribution comes from the call options and for the worst performing outcomes from the put options. 
 
\begin{figure}[htp]
\centering
\begin{tabular}{c}
          \includegraphics[width=100mm]{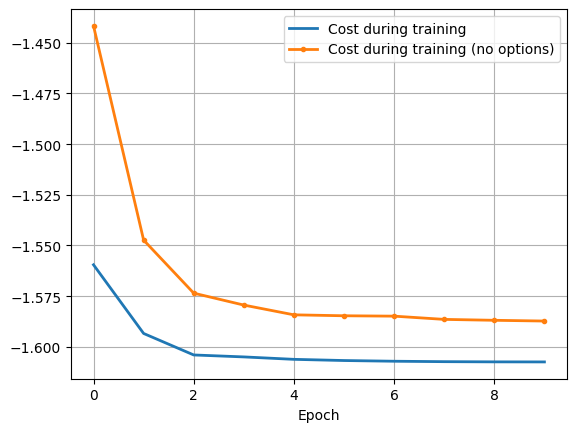}
          \end{tabular}
\caption{Convergence of the loss functions for D-TIPO in blue and D-TIP in orange as a function of the number of training epochs.}\label{cost}
\end{figure}

\begin{figure}[htp]
\centering
\begin{tabular}{cc}
     \includegraphics[width=81mm]{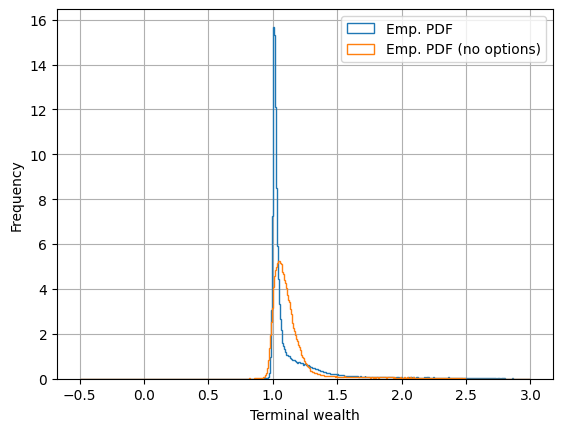}&
          \includegraphics[width=81mm]{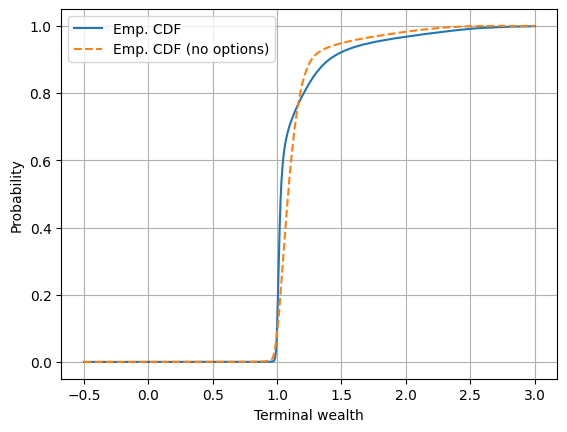}\\
          \includegraphics[width=81mm]{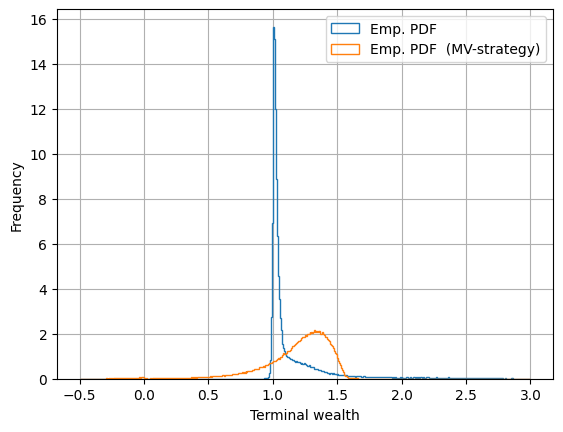}&  \includegraphics[width=81mm]{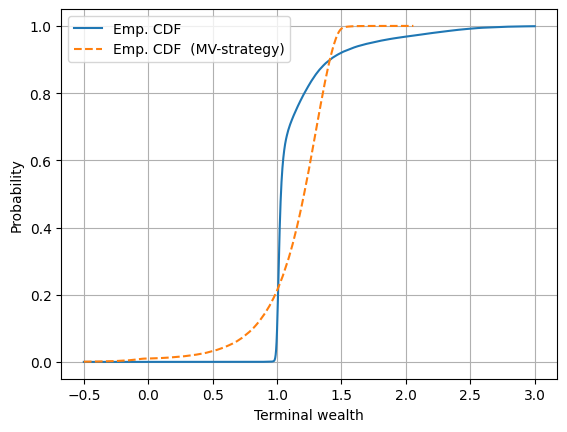}
\end{tabular}
\caption{Comparison of empirical pdfs (left) and CDFs (right) for the D-TIPO and D-TIP strategies (top) and the MV-strategy (bottom).}\label{opts_no_opts_compare}
\end{figure}

 In Section \ref{NR_MV}, we have a closed-form solution as a reference, which made it easy to evaluate the algorithm. Unfortunately, we do not have such a reference in this section. We therefore use two different allocation strategies to compare our results with (for a problem with the same asset process and trading costs): \begin{enumerate}
     \item The same algorithm but with a portfolio consisting of only stocks and a bond, \textit{i.e.,} D-TIP;
     \item The MV-strategy from Section \ref{NR_MV} which generates the same mean as D-TIPO does (1.14). This means that we compare with a strategy that is optimal for classical MV-optimization problem and apply it to the setting in this section.
 \end{enumerate}
For the first comparison, we use the same objective function, but a more restrictive investment directive in the sense that we are not allowed to allocate into the options. Therefore, a reasonable comparison is the optimal (up to the optimization algorithm) value of objective function. If the objective function converges to the same value during optimization with and without options in the portfolio, this would imply that it is not beneficial to add options to the portfolio, \textit{i.e.,} D-TIPO performs in line with D-TIP. If, on the other hand, it converges to a higher value with options in the portfolio, we conclude that it adds value to include options in the portfolio, \textit{i.e.,} D-TIPO performs better than D-TIP. In Figure \ref{cost}, it is clear that the loss function converges to a lower value when options are allowed (we are here using the machine learning convention of a loss function, which is defined by multiplying the objective function by -1).
In Figure \ref{opts_no_opts_compare}, top figures, the allocation strategies with and without options are compared in terms of the distribution of the terminal wealth. From left to right, the empirical pdfs and the empirical CDFs are compared. We note that with options we observe \textit{i}) a thinner left tail, \textit{ii}) a higher density around the expected terminal wealth, and \textit{iii}) a fatter right tail. At least the first and the last features are beneficial since the objective function aims to prevent large losses (by the lower expected shortfall term) and encourage large gains (by the upper expected shortfall term). The reason for this is the additional flexibility to shape the distribution, which is offered by adding options to the portfolio.

For the second comparison, in which we compare the D-TIPO strategy with the MV-strategy, it is no longer suitable to compare the objective functions. The reason for this is that the MV-strategy is obtained as the optimal allocation strategy for a different problem, with other asset dynamics as well as another objective function. The best we can do is to compare the distributions of the terminal wealth, which is also done in Figure \ref{opts_no_opts_compare}. We note the same differences as in the comparison above which is not surprising since the MV-strategy only allocates into stocks and the bond. More interesting is that, in contrast to our strategies both with and without options, we encounter a fatter left tail than the right tail, which is non-desirable from a practical perspective. The reason for this is the limitation of the MV-objective function distinguish between downside risk and upside potential, \textit{i.e.,} it does not only penalize downside risk, but also upside potential.

Similar conclusions can be drawn from Table \ref{tab:2}, in which the empirical mean values, variances and expected shortfalls of the terminal wealth from the three algorithms are compared. For all measures but the variance the portfolio with options performs best. The reason for the variance to be lower without options is simply because the right tail, with a possibility of very high returns. From a practical perspective, this is clearly not a problem. In the rightmost column of Table \ref{tab:2} the transaction cost as a percentage of the initial wealth is expressed. By using the D-TIPO strategy, we see that the transaction cost is decreased by more than 60\% compared to the D-TIP strategy and by more than 90\% compared to the MV-strategy. This is beneficial not only from the perspective of lower transaction costs but also since less aggressive re-allocation is desirable for most investors.
\begin{table}[]
\begin{center}
\begin{tabular}{l|lllll|l}
 & $\E[R]$ & $\text{Var}[R]$ & $\text{ES}^-_{ p_1}(R)$ & $\text{ES}^+_{ p_2}(R)$ & $U(\theta^*)$ & Trading cost\\ \hline
\text{Strategy with options} &\cellcolor{Gray}1.146 & 0.0806 &\cellcolor{Gray}0.971 & \cellcolor{Gray}2.176 & \cellcolor{Gray}1.609 &\cellcolor{Gray}0.386\%\\
\text{Strategy without options} & 1.140 &\cellcolor{Gray}0.0449 & 0.931 & 1.933 & 1.583 & 1.01\%\\
\text{Mean-variance strategy} &\cellcolor{Gray}1.146 & 0.0769 & -0.208 & 1.479 & 1.209 & 3.98\%
\\ \hline
\end{tabular}
\end{center}
\caption{Comparison of our three strategies in terms of empirical mean values, variances, expected shortfalls and the full loss function. Note that for the MV-strategy, the risk parameter $\lambda$, is set to make the mean coincide with the mean obtained from the strategy with options. The trading cost is calculated as a percentage of the initial wealth and reflects how volatile the portfolio re-allocations are.}
\label{tab:2}
\end{table}

\subsubsection{Testing for robustness}
In this paper, we model the market with a jump-diffusion SDE, but as already stated, any model that can generate enough samples would fit the framework. On the other hand, whatever model is used, the only thing that we can be sure of is that real world market does not behave exactly as the model. Therefore, we aim to test the robustness of the algorithms for model miss-specification. This is carried out by applying the strategies from above with either significantly higher or significantly lower volatility of the underlying asset process. In the high volatility case, we multiply $\sigma$ from \eqref{parameter_values} by two and in the low volatility case, we divide $\sigma$ by two.  

In Table \ref{tab:3}, we see that the D-TIPO and D-TIP strategies significantly outperform the MV-strategy both with increased and decreased volatility. This is due to the fact that the volatility is a sensitive parameter in the closed-form strategy implying that if the volatility is either larger or smaller, then the strategy deviates significantly from optimality.\footnote{Bear in mind that for the MV-strategy optimality is only in the sense of mean and variance since the expected shortfall terms do not enter the objective function.} Most notable is the lower expected shortfall, which expresses a loss of $172.8\%$ and $98.4\%$ and the trading costs at $10.1\%$ and $2.74\%$ for the higher and lower volatilities, respectively. In the comparison between the strategies, it does not come as a surprise that options in the portfolio are particularly beneficial when the volatility increases and are less beneficial when the volatility decreases. Moreover, we note that the variance is larger for the strategy with options than without options, especially in the case with higher volatility. As described above, this is due to the fatter right tail of the distribution of terminal wealth and should not be viewed as a problem.
\begin{table}[]
\begin{center}
\begin{tabular}{l|lllll|l}
 & $\E[R]$ & $\text{Var}[R]$ & $\text{ES}^-_{ p_1}(R)$ & $\text{ES}^+_{ p_2}(R)$ & $U(\theta^*)$ & Trading cost\\ \hline
        \multicolumn{7}{c}{Evaluation with increased volatility for the underlying assets ($\sigma\mapsto 2\times\sigma$).}
\\ \hline
\text{Strategy with options} & 1.318 & 0.660 &\cellcolor{Gray} 0.763 &\cellcolor{Gray}3.268 &\cellcolor{Gray}1.540 & \cellcolor{Gray}0.838\%\\
\text{Strategy without options} & \cellcolor{Gray}1.350 &\cellcolor{Gray} 0.175 & 0.734 & 2.635 & 1.532 & 1.23\%\\
\text{Mean-variance strategy} & 1.081 & 0.674 & -0.728 & 1.460 & 0.668 & 10.1\%
\\ \hline
        \multicolumn{7}{c}{Evaluation with decreased volatility for the underlying assets ($\sigma\mapsto 0.5\times\sigma$).}
\\ \hline
\text{Strategy with options} & 1.074 & 0.0262 & \cellcolor{Gray}0.969 & \cellcolor{Gray}1.620 & 1.506 & \cellcolor{Gray}0.261\%\\
\text{Strategy without options} & 1.143 & \cellcolor{Gray}0.0160 & 0.957 & 1.548 & \cellcolor{Gray}1.569 & 0.636\%\\
\text{Mean-variance strategy} & \cellcolor{Gray}1.163 & 0.0569 & 0.0160 & 1.487 & 1.301 & 2.74\%
\end{tabular}
\end{center}
\caption{Comparison of our three strategies in terms of empirical mean values, variances, expected shortfalls and the full loss function. Note that for the MV-strategy, the risk parameter $\lambda$, is set to make the mean coincide with the mean obtained from the D-TIPO strategy. The trading cost is calculated as a percentage of the initial wealth and reflects how volatile the portfolio re-allocations are.}
\label{tab:3}
\end{table}

\section{Conclusions}\label{conclusions}
The main conclusions drawn from this paper can be summarized in three parts: 1) It is important to be careful with the choice of objective function, in order to reflect the true incentives of a rational trader; 2) Adding options makes shaping the distribution of the terminal wealth more flexible, due to the asymmetric distribution of option returns. Moreover, adding options may significantly reduce the re-allocation and in turn the trading cost; 3) A sequence of neural networks can produce a high quality allocation strategy in high dimensions (many assets, options and also approximating strike prices for each option). 

The extension to trading options in a dynamic setting is straightforward, as long as we have access to an efficient method for option valuation along stochastic asset trajectories. To approximate other options parameters, such as terminal time etc., is a straightforward extension. 

\section*{Acknowledgments}
We acknowledge the funding of our research by the European Union, under the H2020-EU.1.3.1. MSCA-ITN-2018 scheme, Grant 813261. The authors would like to thank Leonardo Perotti for insightful comments which significantly improved this manuscript.

\end{document}